\documentclass[sigconf,nonacm]{acmart}
\usepackage{listings}
\usepackage{xcolor}
\usepackage[utf8]{inputenc}
\usepackage{array}
\usepackage{wrapfig}
\usepackage{multirow}
\usepackage{tabularx}
\usepackage[draft=true]{minted}

\definecolor{codegreen}{rgb}{0,0.6,0}
\definecolor{codegray}{rgb}{0.5,0.5,0.5}
\definecolor{codepurple}{rgb}{0.58,0,0.82}
\definecolor{backcolour}{rgb}{0.95,0.95,0.92}

\lstdefinestyle{mystyle}{
    backgroundcolor=\color{backcolour},   
    commentstyle=\color{codegreen},
    keywordstyle=\color{magenta},
    numberstyle=\tiny\color{codegray},
    stringstyle=\color{codepurple},
    basicstyle=\ttfamily\footnotesize,
    breakatwhitespace=false,         
    breaklines=true,                 
    captionpos=b,                    
    keepspaces=true,                 
    numbers=left,                    
    numbersep=5pt,                  
    showspaces=false,                
    showstringspaces=false,
    showtabs=false,                  
    tabsize=2
}
\colorlet{punct}{red!60!black}
\definecolor{background}{HTML}{EEEEEE}
\definecolor{delim}{RGB}{20,105,176}
\colorlet{numb}{magenta!60!black}

\lstdefinelanguage{json}{
    basicstyle=\normalfont\ttfamily,
    numbers=left,
    numberstyle=\scriptsize,
    stepnumber=1,
    numbersep=8pt,
    showstringspaces=false,
    breaklines=true,
    frame=lines,
    backgroundcolor=\color{background},
    literate=
     *{0}{{{\color{numb}0}}}{1}
      {1}{{{\color{numb}1}}}{1}
      {2}{{{\color{numb}2}}}{1}
      {3}{{{\color{numb}3}}}{1}
      {4}{{{\color{numb}4}}}{1}
      {5}{{{\color{numb}5}}}{1}
      {6}{{{\color{numb}6}}}{1}
      {7}{{{\color{numb}7}}}{1}
      {8}{{{\color{numb}8}}}{1}
      {9}{{{\color{numb}9}}}{1}
      {:}{{{\color{punct}{:}}}}{1}
      {,}{{{\color{punct}{,}}}}{1}
      {\{}{{{\color{delim}{\{}}}}{1}
      {\}}{{{\color{delim}{\}}}}}{1}
      {[}{{{\color{delim}{[}}}}{1}
      {]}{{{\color{delim}{]}}}}{1},
}

\AtBeginDocument{%
  \providecommand\BibTeX{{%
    \normalfont B\kern-0.5em{\scshape i\kern-0.25em b}\kern-0.8em\TeX}}}

\setcopyright{acmcopyright}
\copyrightyear{2020}
\acmYear{2020}
\acmDOI{10.1145/1122445.1122456}

\acmConference[SIGMOD '21]{SIGMOD '21: ACM SIGMOD/PODS International Conference on Management of Data}{June 20--25, 2021}{Xi'an, Shaanxi, China}
\acmBooktitle{SIGMOD '21: ACM SIGMOD/PODS International Conference on Management of Data, June 20--25, 2021, Xi'an, Shaanxi, China}
\acmPrice{15.00}
\acmISBN{978-1-4503-XXXX-X/18/06}

\begin{document}

\title{Apache Submarine: A Unified Machine Learning Platform Made Simple}


\author{Kai-Hsun Chen}
\affiliation{%
  \institution{Academia Sinica}}
\email{kaihsun@apache.org}

\author{Huan-Ping Su}
\affiliation{%
  \institution{Academia Sinica}}
\email{pingsutw@apache.org}

\author{Wei-Chiu Chuang}
\affiliation{%
  \institution{Cloudera}}
\email{weichiu@cloudera.com}

\author{Hung-Chang Hsiao}
\affiliation{%
  \institution{National Cheng Kung University}
}
\email{hchsiao@csie.ncku.edu.tw}

\author{Wangda Tan}
\affiliation{%
  \institution{Cloudera}}
\email{wtan@cloudera.com}

\author{Zhankun Tang}
\affiliation{%
  \institution{Cloudera}
}
\email{ztang@cloudera.com}

\author{Xun Liu}
\affiliation{%
  \institution{DiDi}
}
\email{xunliu@didichuxing.com}

\author{Yanbo Liang}
\affiliation{%
  \institution{Facebook}
}
\email{ybliang@fb.com}

\author{Wen-Chih Lo}
\affiliation{%
  \institution{National Cheng Kung University}
}
\email{wenchih@apache.org}

\author{Wanqiang Ji}
\affiliation{%
  \institution{JD.com}
}
\email{jiwanqiang@jd.com}

\author{Byron Hsu}
\affiliation{%
  \institution{UC Berkeley}
}
\email{byronhsu@apache.org}

\author{Keqiu Hu}
\affiliation{%
  \institution{LinkedIn}
}
\email{Khu@linkedin.com}

\author{HuiYang Jian}
\affiliation{%
  \institution{KE Holdings}
}
\email{jianhuiyang001@ke.com}

\author{Quan Zhou}
\affiliation{%
  \institution{Ant Group}
}
\email{qiyuan.zq@antgroup.com}

\author{Chien-Min Wang}
\affiliation{%
  \institution{Academia Sinica}
}
\email{cmwang@iis.sinica.edu.tw}

\renewcommand{\shortauthors}{Chen and Su, et al.}

\begin{abstract}
As machine learning is applied more widely, it is necessary to have a machine learning platform for both infrastructure administrators and users including expert data scientists and citizen data scientists \cite{CitizenDataScientist} to improve their productivity. However, existing machine learning platforms are ill-equipped to address the “Machine Learning tech debts” \cite{sculley2015hidden} such as glue code, reproducibility, and portability. Furthermore, existing platforms only take expert data scientists into consideration, and thus they are inflexible for infrastructure administrators and non-user-friendly for citizen data scientists. We propose Submarine, a unified machine learning platform, to address the challenges. 


For infrastructure administrators, Submarine is a flexible and portable platform. To improve resource efficiency and ensure portability, Submarine supports both on-premise clusters including Kubernetes \cite{Kubernetes} and Hadoop YARN \cite{YARN}, and cloud services, rather than only focusing on Kubernetes. In addition, Submarine supports multiple machine learning frameworks to avoid tight coupling between the modeling and infrastructure layers caused by glue code.

For expert data scientists, Submarine is a unified system to streamline the workflow from idea to serving models in production. Data scientists without experience in distributed systems can operate Submarine smoothly, since Submarine hides the details of infrastructure.

For citizen data scientists, Submarine is an easy-to-use platform to save their time on machine learning framework API and help them focus on domain-specific problems. To elaborate, Submarine provides a high-level SDK that allows users to develop their models in a few lines of code. Also, the Submarine Predefined Template Service enables users to run experiments without writing a single line of code.

Submarine has been widely used in many technology companies, including Ke.com and LinkedIn. We present two use cases in Section~\ref{section:use_cases}.
\end{abstract}


\maketitle
\pagestyle{empty}

\section{Introduction \label{section:introduction}}
The ubiquity of machine learning in all aspects of businesses imposes plenty of challenges for infrastructure administrators, expert data scientists, and citizen data scientists. For example, as described in \cite{krizhevsky2014weird}, models with large parameters' space cannot fit into the memory of a single machine, necessitating model parallelism support. In addition, data parallelism support is necessary for models with high computation cost to shorten the training process. Therefore, distributed training techniques like data parallelism and model parallelism which leverage distributed computing resources to train the models are necessary.

Despite rapid development of distributed machine learning platforms, existing platforms fail to meet the needs of both infrastructure administrators and users, who have different requirements for the system. We will discuss the problems in the existing platforms respectively from perspectives of both infrastructure administrators and users including expert data scientists and non-expert users, or “citizen data scientists” \cite{CitizenDataScientist}, who are experts in other fields such as finance, medicine, and education, but who are not necessarily familiar with machine learning techniques.

From an expert data scientists’ perspective, a machine learning platform must be able to help them improve the productivity, shortening the model development time to production. However, existing platforms are not properly designed to meet this expectation. Firstly, existing platforms like TFX \cite{TFX} do not cover the whole model lifecycle which is composed mainly of data preparation, prototyping, model training, and model serving, forcing data scientists to switch toolsets for different stages of the model development, creating an inconsistent user experience. Secondly, developing on these platforms requires deep expertise. For instance, MLflow \cite{Mlflow} requires data scientists to write a Kubernetes Job Spec to run an MLflow project on Kubernetes. In Determined \cite{Determined}, data scientists need to rewrite their model code to integrate with its API. 

From a citizen data scientists’ perspective, they spend most of their time understanding the business problems and are less interested in learning machine learning APIs. Hence, citizen data scientists expect a machine learning platform which saves their time on machine learning framework API and helps them focus on domain-specific problems. However, existing platforms only take expert data scientists into consideration, and thus they are non-user-friendly for citizen data scientists. Ironically, the citizen data scientist population is a much larger and fast growing community than the expert data scientist population. A new platform that considers both expert data scientists and citizen data scientists is a must.

From an infrastructure administrators’ perspective, a machine learning platform should ideally be resource-efficient and flexible, but existing platforms were not capable of achieving these goals. Firstly, most platforms such as Kubeflow \cite{Kubeflow} only support Kubernetes as a resource orchestrator. However, many technology companies like LinkedIn have deeply invested in other orchestration technologies, such as YARN. Secondly, some platforms like Metaflow \cite{Metaflow} are built on top of proprietary services offered by public cloud providers such as Amazon Web Services, Microsoft Azure, or Google Cloud Platform. Hence, they are not suitable for companies that deploy machine learning platforms in data centers. Thirdly, existing platforms like TFX are tightly integrated with a single machine learning framework (e.g. TensorFlow \cite{Tensorflow}). The tight coupling between the modeling layer and the platform layer limits the flexibility and extensibility of the system, because the developers lose the ability to use a different machine learning framework, which may be more efficient for a different use case.

We made a decision to build Apache Submarine, a unified machine learning platform, to address the problems mentioned above. A “unified” machine learning platform describes a platform that provides a unified solution for infrastructure administrators, expert data scientists and citizen data scientists. Submarine takes both infrastructure administrators and users including expert data scientists and citizen data scientists into consideration. For expert data scientists, Submarine is a unified platform that provides data scientists with a consistent set of tools built for different stages in the model development lifecycle. In addition, data scientists can operate Submarine without extra infrastructure knowledge. For citizen data scientists, Submarine provides a high-level SDK that allows users to develop their models in a few lines of Python code. We interviewed data scientists and found that the machine learning algorithms, such as the prediction of click-through rate, used in most applications can be modeled after a few parameterized templates, and a number of templates are packaged into the Submarine Predefined Template Service to enable citizen data scientists to run experiments without code. Additionally, all components in Submarine are well integrated with Submarine Workbench, and thus citizen data scientists can operate the platform easily. For infrastructure administrators, Submarine supports both on-premise clusters managed by Kubernetes and YARN and clouds to ensure portability and resource-efficiency. Also, Submarine supports popular machine learning frameworks including TensorFlow, PyTorch \cite{PyTorch}, and MXNet \cite{MXNet} to ensure the flexibility and extensibility of the system.

Machine Learning Platform is a new and exciting research topic, where it sits at the intersection of Systems and Machine Learning. Therefore, there are not many best practices available so far. Leading technology companies \cite{SpotifyMachineLearningInfrastructure} tried to build their in-house machine learning platforms, often to fail or only to claim success after repeated trials. We decided to develop Submarine as an open source project governed by the Apache Software Foundation so we can build a community and share the best practices. Today, the project is joined by contributors all over the world, including the developers from Cloudera, DiDi, Facebook, JD.com, LinkedIn, KE Holdings, Ant Group, Microsoft, among others.

The rest of the paper is organized as follows. At first, Section~\ref{section:related_work} gives a brief overview of existing machine learning platforms. Section~\ref{section:architecture} and Section~\ref{section:in_progress_work} describe working and in-progress components in Submarine. Then, we discuss the uniqueness of Submarine and compare Submarine with other open source platforms in Section~\ref{section:discussion}. We present two use cases in technology companies including LinkedIn and Ke.com in Section~\ref{section:use_cases} and discuss future works in Section~\ref{section:future_work}. Finally, we conclude this work in Section~\ref{section:conclusion}. 

\section{Related work \label{section:related_work}}
Nowadays, many machine learning platforms have been developed by companies and open source communities. 
We briefly overview some major open source works including Google’s TFX \cite{TFX}, Kubeflow \cite{Kubeflow}, Netflix’s Metaflow \cite{Metaflow}, Determined-AI’s Determined \cite{Determined}, Microsoft’s NNI (Neural Network Intelligence) \cite{NNI}, Tencent’s Angel-ML \cite{Angelml}, Databricks' MLflow \cite{Mlflow}, and Anyscale's Ray \cite{Ray} in this section.

TFX is a toolkit for building machine learning pipelines. 
Users can define their TFX pipelines which are a sequence of components from data engineering to model serving. 
Moreover, with TFX pipelines, users can orchestrate their machine learning workflows on several platforms, such as
Apache Airflow \cite{Airflow} Apache Beam \cite{Beam}, and Kubeflow Pipelines.

Kubeflow is an end-to-end Machine Learning platform which aims to make deployments of machine learning workflows on Kubernetes. 
Kubeflow also provides an interface that allows users to track and compare experiments. 
Thus, users can decide which experiment is the best and use it as the main source for future steps.

Metaflow is a Python library that helps scientists and engineers build and manage real-life data science projects.
Metaflow provides a unified API to access the infrastructure stack that is required to execute data science projects, from prototype to production.

Determined is an open source deep learning training platform that makes building models fast and easy. 
Determined allows users to develop and train models in customizable containers, which enable simple and consistent dependency management throughout the model development lifecycle.

NNI (Neural Network Intelligence) developed by Microsoft is a toolkit to automate feature engineering, neural architecture search, hyperparameter tuning, and model compression.
The tool manages automated machine learning (AutoML) experiments and runs experiments' trial jobs generated by tuning algorithms to search for the best neural architecture and hyper-parameters.

Angel-ML is developed by Tencent. 
It is a high-performance distributed machine learning platform based on Parameter Server~\cite{smola2010architecture} and provides multiple click-through rate (CTR) algorithms written by Java and Scala.

MLflow is an open source Machine Learning platform developed by Databricks. 
With MLflow, users can easily manage the machine learning lifecycle, including tracking experiments, packaging code into reproducible runs, and sharing and deploying models.

Ray provides a unified API for distributed applications. It supports high-efficient stateful and stateless computation to allow a wide variety of scenarios (AutoML, RL, distributed training, etc.) built on top of it.

Several in-house machine learning platforms are used to streamline the workflow from idea to serving models in production, such as Facebook’s FBLearner Flow \cite{Fblearner}, Twitter's Cortex \cite{TwitterCortex}, Uber’s Michelangelo \cite{Michelangelo}, and Airbnb’s Bighead \cite{Bighead}. 
However, these platforms are not yet open sourced.

Also, cloud vendors provide services including training, serving, and model management such as Amazon’s SageMaker \cite{Sagemaker}, Microsoft Azure Machine Learning \cite{AzureML}, Google's Vertex AI \cite{VertexAI}, Cloudera Data Science Workbench \cite{Cloudera}, and Valohai MLOps platform \cite{Valohai}. Like the in-house machine learning platforms, they are proprietary.



\section{Architecture \label{section:architecture}}

\begin{figure}[h]
  \centering
  \includegraphics[width=\linewidth]{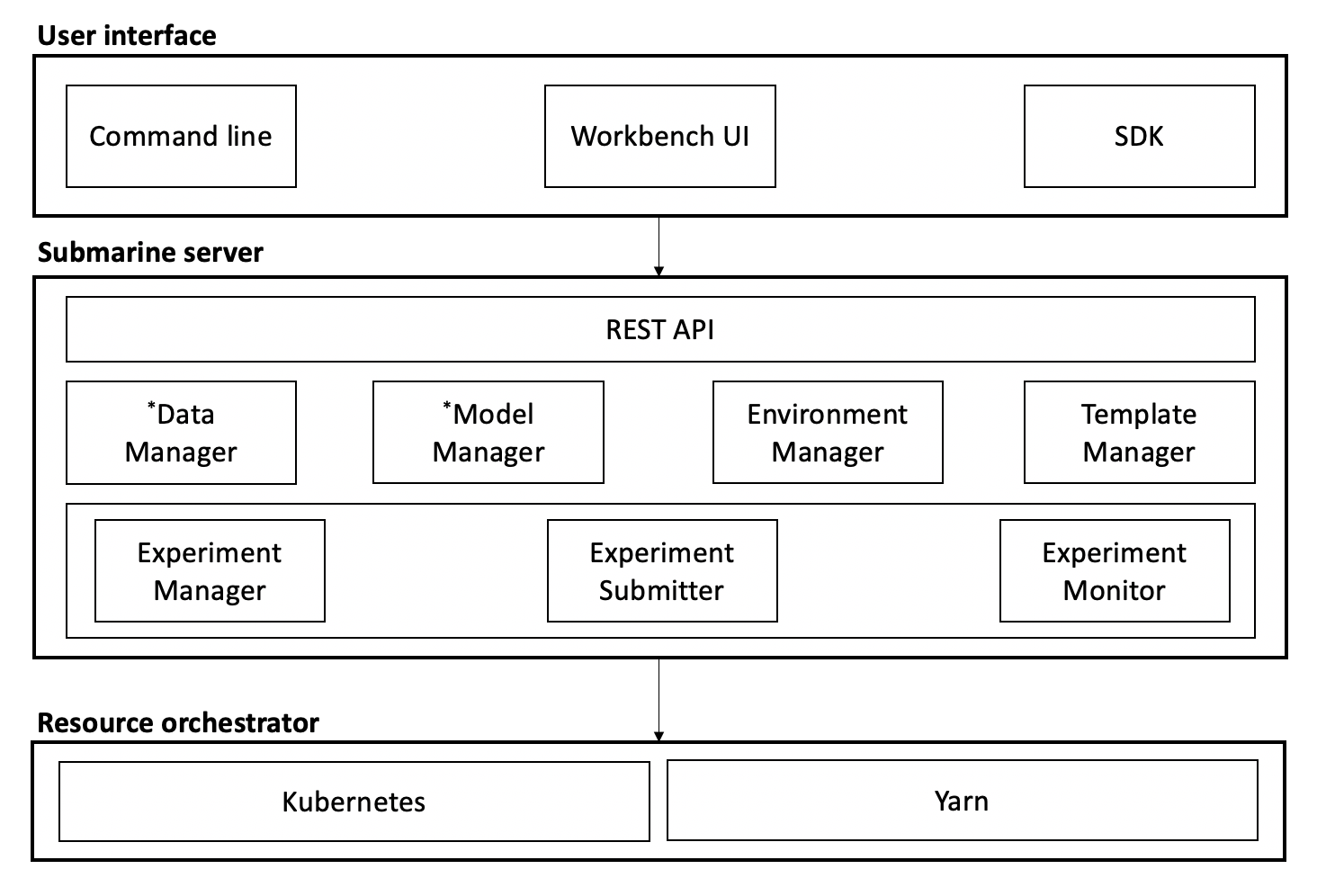}
  \caption{Submarine architecture overview}
  \label{fig:submarine_architecture}
\end{figure}

As shown in Fig.~\ref{fig:submarine_architecture}, Submarine consists of three parts including user interface, Submarine server, and resource orchestrator. We will discuss each part as follows. Moreover, the components marked with asterisks are in-progress works or future works, and in-progress components are discussed in Section~\ref{section:in_progress_work}.

\subsection{Submarine user interface}
Submarine provides three types of user interface, including command line (CLI), software development kit (SDK), and workbench UI. To elaborate, these user interfaces manipulate each component in the model lifecycle via REST API exposed by Submarine server. The REST API service handles HTTP requests and is responsible for authentication.

\subsubsection{Submarine CLI}
Users can interact with each component in Submarine server through command line. Take Listing 1 as an example, this command is to run a MNIST \cite{lecun2010mnist} training experiment, a distributed training job, on Submarine.

\begin{minted}[breaklines, frame=single,framesep=5pt]{bash}
submarine job run 
 --name mnist \
 --framework TensorFlow
 --num_workers 4 \
 --worker_resources memory=4G,gpu=4,vcores=4 \
 --num_ps 1 \
 --ps_resources memory=2G,vcores=2 \
 --worker_launch_cmd "python mnist.py" \
 --ps_launch_cmd "python mnist.py" \
 --insecure \
 --conf tony.containers.resources=mnist.py 
\end{minted}
\captionof{listing}{A command line MNIST training example}

\subsubsection{Submarine Python SDK}
Users can interact with each component in Submarine server through a Python SDK. For instance, like Listing 1, the example in Listing 2 deploys a MNIST experiment on Submarine via Python SDK too. In addition, the Python SDK can also be used in the notebook service in workbench. 

~\\
\begin{minted}[breaklines, frame=single,framesep=5pt]{python}

submarine_client = submarine.ExperimentClient()
environment = EnvironmentSpec(
                image='submarine:tf-mnist')
experiment_meta = ExperimentMeta(
                name='mnist',
                namespace='default',
                framework='TensorFlow',
                cmd='python mnist.py')
                
ps_spec = ExperimentTaskSpec(
                resources='cpu=2,
                          memory=2G,
                          replicas=1')
                
worker_spec = ExperimentTaskSpec(
                resources='cpu=4,gpu=4,
                          memory=4G,
                          replicas=4')
experiment_spec = ExperimentSpec(
                meta=experiment_meta,
                environment=environment,
                spec={'Ps': ps_spec,
                      'Worker': worker_spec})

\end{minted}
\captionof{listing}{An SDK example}
~\\
Furthermore, Submarine Python SDK is an intuitive high-level SDK to streamline the workflow from idea to serving models in production. Specifically, with the SDK, data scientists can develop their models in a few lines of Python code instead of building models from scratch. For instance, the prediction of click-through rate (CTR) is critical in recommender systems, and DeepFM \cite{DeepFM} is one of the most popular models related to CTR prediction. As shown in Listing 3, users can build a DeepFM model in just four lines of Python code.

~\\
\begin{minted}[breaklines, frame=single,framesep=5pt]{python}
from submarine.ml.tensorflow.model import DeepFM
model = DeepFM(json_path=deepfm.json)
model.train()
result = model.evaluate()
print("Model AUC : ", result)
\end{minted}
\captionof{listing}{A high-level SDK example}

\subsubsection{Workbench}
\begin{figure}[htbp]
  \centering
  \includegraphics[width=\linewidth]{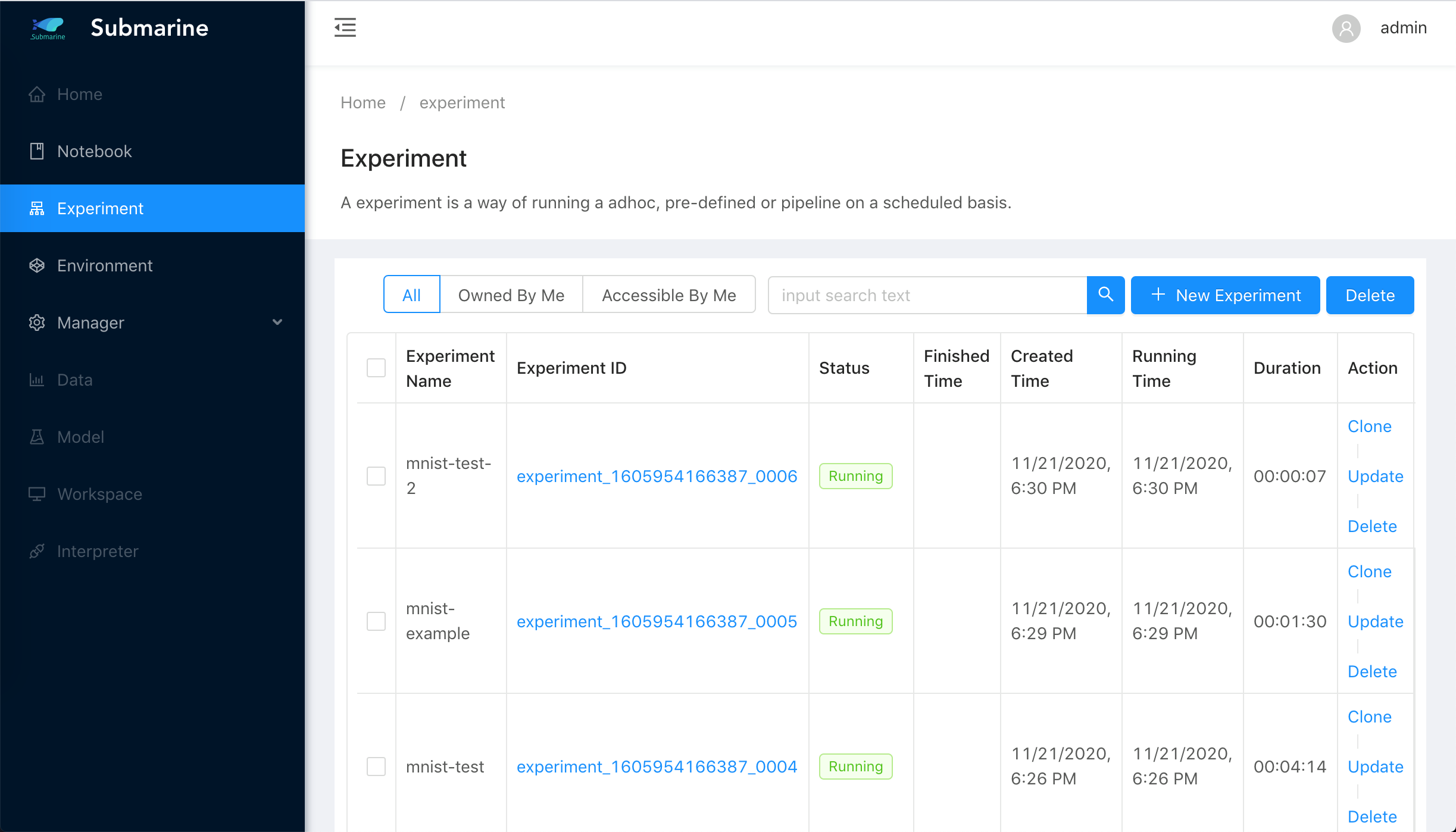}
  \caption{A screenshot of Submarine Workbench}
  \label{fig:submarine_workbench}
\end{figure}

To provide better user experiences, Submarine comes with an easy-to-use workbench UI for users to manage each stage in the model lifecycle. See Fig.~\ref{fig:submarine_workbench} for a screenshot of the workbench UI. We will explain how workbench interacts with each stage in the model lifecycle.

\begin{itemize}
\item {\textbf {Data preparation (future work)}}: Users can select features existing in feature stores instead of rebuilding these features repeatedly via Submarine Workbench. Also, Submarine plans to visualize training data in a meaningful way which makes it easier to identify patterns and trends than looking through thousands of rows on a spreadsheet.

\item{\textbf {Prototyping}}: Submarine provides Jupyter notebook \cite{Jupyter}, a user-defined prototyping environment, in workbench. Users can develop their machine learning algorithms with popular machine learning frameworks or Submarine high-level Python SDK.
\item{\textbf {Model training}}: After model development is finished, users can easily deploy experiments, distributed training jobs, on Kubernetes or YARN through the workbench UI. Moreover, Submarine provides metric visualization of experiments in an innovative way which enables data scientists to compare the performance of experiments easily.
\item{\textbf {Model serving (future work)}}: After data scientists select the models, users can serve the trained models to production, and perform online/offline inference through the workbench UI. 
\end{itemize}

\subsection{Submarine server}
Submarine server plays the role of control plane in Submarine. Submarine server has two main functionalities. Firstly, Submarine server exposes a REST API for users to manipulate each component in the model lifecycle. Secondly, Submarine server consists of several core services, including data manager, model manager, environment manager, template manager, experiment manager, experiment submitter, and experiment monitor.

\subsubsection{Submarine Environment Service}
It is difficult for data scientists to write Dockerfiles and manage Docker images. Therefore, Submarine encapsulates Docker and Virtual Machine including VirtualBox images, VMWare images, Amazon Machine Images and Azure Virtual Machine images in Submarine Environment Service. Submarine Environment Service provides a convenient way for data scientists to package up applications and preconfigured library dependencies. An environment consists of base libraries such as operating systems, CUDA and GPU drivers, and library dependencies such as Python and TensorFlow. With these environments, Jupyter notebooks become reusable and experiments become reproducible.

Furthermore, we select Conda as our dependency management system, and thus users can easily add/remove their dependencies for different languages such as Python, R, Ruby, Lua, Scala, Java, JavaScript, C/ C++, FORTRAN, and more. In addition, users can also define an environment via a YAML file.

\subsubsection{Submarine Experiment Service}
\begin{figure}[h]
  \centering
  \includegraphics[width=\linewidth]{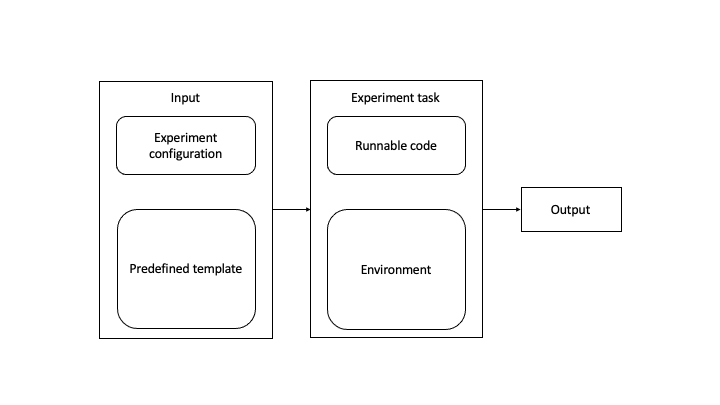}
  \caption{Submarine experiment}
  \label{fig:submarine_experiment}
\end{figure}

Submarine provides a flexible and easy-to-use abstraction, Submarine experiment, to help users achieve better user experiences. As shown in Fig.~\ref{fig:submarine_experiment}, a Submarine experiment consists of three parts: \textit{Input}, \textit{Submarine experiment task}, and \textit{Output}.

\begin{itemize}
\item {\textbf {Input}}: Input consists of two components as shown in Fig.~\ref{fig:submarine_experiment}. Experiment configuration specifies resource constraints, training data, Submarine environment, placement constraints, number of experiment workers, and so on. Moreover, the predefined template is an important feature which enables users to run experiments easily, and it is an optional parameter. More details about the Predefined Template Service is described in Section ~\ref{ssse:predefined_template_service}.
\item {\textbf {Experiment task}}: In Fig.~\ref{fig:submarine_experiment}, runnable code such as Python scripts can be used to train a model or to process data. Moreover, the environment specified in the experiment configuration includes the experiment dependencies such as Python and TensorFlow and OS/Base libraries such as Ubuntu and CUDA. With the environment, experiments become reproducible.
\item {\textbf {Output}}: The main output of the experiment is artifacts which may include models. Furthermore, logs and metrics are used to troubleshoot bugs and evaluate the quality of models. In addition, metric visualization is provided in Submarine Workbench.
\end{itemize}

\begin{figure}[h]
  \centering
  \includegraphics[width=\linewidth]{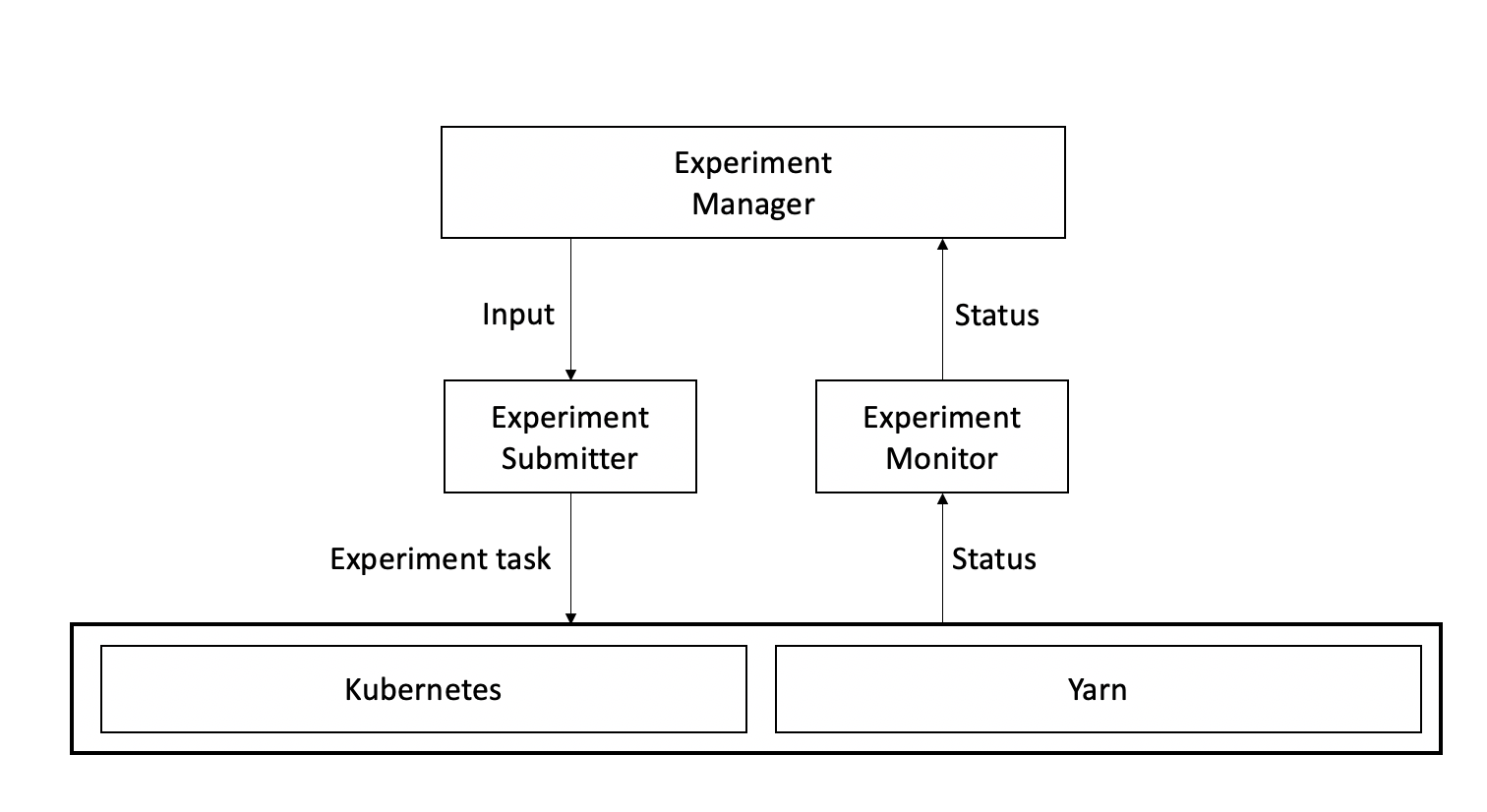}
  \caption{Architecture of Submarine Experiment Service}
  \label{fig:submarine_experiment_architecture}
\end{figure}

As shown in Fig.~\ref{fig:submarine_experiment_architecture}, Submarine implements the experiment service via three components: experiment manager, experiment submitter and experiment monitor. These components will be discussed as follows.

\begin{itemize}
\item {\textbf{Experiment manager}}: Experiment manager listens to experiment-related requests from users. When the experiment manager accepts a request, it persists the experiment metadata in a database so that experiments become easy to compare and reproducible. In addition, the experiment manager will forward the request to the experiment submitter.

\item {\textbf{Experiment submitter}}: The experiments can be launched in YARN cluster, Kubernetes cluster or locally. Hence, Submarine provides two types of submitters, YARN submitter and Kubernetes submitter, to submit the experiment to the cluster managed by YARN or Kubernetes. Furthermore, YARN submitter uses Tensorflow on YARN (TonY) \cite{Tony} as the runtime to run experiments. In addition, the Kubernetes submitter used operators such as tf-operator as the runtime. To ensure extensibility, Submarine provides a submitter abstraction, and thus users can implement tailor-made submitters to support new container orchestration frameworks such as Docker Swarm \cite{Swarm}.

\item {\textbf{Experiment monitor}}: Experiment monitor tracks the status of experiments and records important events and sends them to the experiment manager. This information plays a key role to predict the success or failure of the in-progress experiment.

\end{itemize}

\subsubsection{Submarine Predefined Template Service \label{ssse:predefined_template_service}}
\begin{figure}[h]
  \centering
  \includegraphics[width=\linewidth]{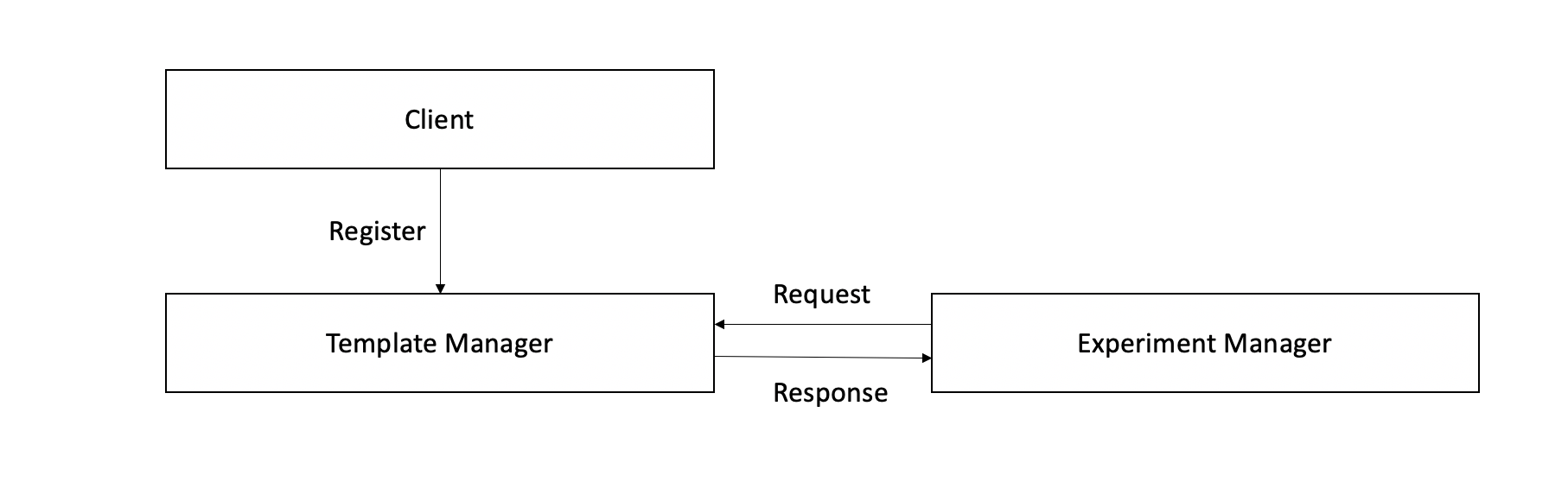}
  \caption{Template manager mechanism}
  \label{fig:submarine_template_manager}
\end{figure}

The real-world data scientists spend the majority of time on hyperparameter tuning and training with different datasets instead of designing new model algorithms. Submarine provides the predefined template abstraction to make users run experiments with different hyperparameters easily. Fig.~\ref{fig:submarine_template_manager} demonstrates the mechanism of the Predefined Template Service which will be discussed as follows.

\hspace{5pt} Firstly, clients define predefined template specifications in JSON files as shown in Listing 4. Furthermore, clients can register these specifications in the template manager for sharing and reusing. Predefined template specifies the experiment specification and model parameters which users need to specify. Take Listing 4 as an example, users just need to specify two parameters, $learning\_rate$ and $batch\_size$, to train a CNN model \cite{CNN} for the MNIST dataset in a distributed manner.

\hspace{5pt} Secondly, in the workbench, citizen data scientists can select the appropriate templates which have been registered in the template manager. With the help of the template, users only need to fill in the parameters which the template requires to run an experiment. In other words, users can run experiments without writing one line of code.

\hspace{5pt} Lastly, the Submarine community has already provided a bunch of templates for popular machine learning applications such as image recognition and click-through rate prediction. For expert data scientists, they can easily register different kinds of templates for different use cases. For citizen data scientists, they can submit an experiment without writing any code.

\begin{figure}[t]
\begin{minted}[breaklines, frame=single,framesep=5pt,fontsize=\small]{JSON}
{
  "name": "tf-mnist-template",
  "author": "Submarine",
  "description": "A template for tf-mnist",
  "parameters": [{
      "name": "learning_rate",
      "value": 0.001,
      "required": true,
    },
    {
      "name": "batch_size",
      "value": 256,
      "required": true,
    }
  ],
  "experimentSpec": {
    "meta": {
      "cmd": "python mnist.py 
      --log_dir=/train/log
      --learning_rate={{learning_rate}}
      --batch_size={{batch_size}}",
      "framework": "TensorFlow",
      "namespace": "default"
    },
    "spec": {
      "Ps": {
        "replicas": 1,
        "resources": "cpu=2,memory=2G"
      },
      "Worker": {
        "replicas": 4,
        "resources": "cpu=4,gpu=4,memory=4G"
      }
    },
    "environment": {
      "image": "submarine:tf-mnist"
    }
  }
}
\end{minted}
\captionof{listing}{A JSON file to define a predefined template specification}
\end{figure}

\subsection{Container orchestration framework}
Submarine supports both Kubernetes and Apache Hadoop YARN as our container orchestration frameworks. By leveraging these frameworks, clusters can scale up/down automatically based on dynamic resource requirements to save costs. In addition, the orchestrators can perform different scheduling strategies to improve resource utilization of clusters.

\section{In-progress work \label{section:in_progress_work}}
\subsection{Automated machine learning (AutoML)}
AutoML is a set of techniques including automated data cleaning, automated feature engineering, hyperparameter optimization, meta learning, and neural architecture search. Because hyperparameter tuning is time-consuming and neural network architecture design is labor-intensive, Submarine plans to support hyperparameter optimization and neural architecture search to shorten the development-to-production lifecycle. With the help of these AutoML algorithms, users can train custom models that are specific to their business needs, with minimum effort and machine learning expertise.

\subsection{Model manager}
A model is a composite of several artifacts, including but not limited to hyperparameters, source code, training sets, environments, and neural network architectures between models.  It is hard for data scientists to track and compare different versions of models. To address the challenges, Submarine plans to provide model management service in model manager to help users manage metadata of models. Models will be versioned to provide reproducibility. Moreover, data scientists can reuse models registered in the model manager to avoid redundant efforts and streamline the workflow from idea to serving models in production.

\section{Discussions \label{section:discussion}}
In Table 1, we compare the major features supported by Apache Submarine and other open source machine learning systems. The notations used in Table 1 are listed in Table 2. In-house platforms such as FBLearner Flow, Michelangelo and Bighead are excluded from this discussion due to their proprietary nature. Detailed descriptions are discussed in the following subsections.

\begin{table*}[htbp]
\vspace*{0.5cm}
\caption{Comparisons among Submarine and other platforms}
\begin{center}
\begin{tabular}{|p{6cm}|c|c|c|c|c|c|c|c|}
\hline
&\textbf{TFX}&\textbf{KF} &\textbf{DT} &\textbf{MF} &\textbf{MLF} &\textbf{NNI} &\textbf{AML} &\textbf{Submarine} \\\hline
Open source                          & v & v & v & v & v & v & v & v \\\hline
Kubernetes                           & v & v & v &   & v & v &   & v \\\hline
YARN                                 &   &   &   &   &   &   & v & v \\\hline
Multi ML frameworks                  &   & v & v & v & v & v & v & v \\\hline
Feature store                        &   & v &   &   &   &   &   & $\triangle$ \\\hline
User-defined prototyping environment &   & v & v &   &   &   &   & v \\\hline
Distributed training                 & v & v & v & v &   & v & v & v \\\hline
High-level training SDK              &   &   &   &   &   &   & v & v \\\hline
Automatic hyperparameter tuning      & v & v & v &   &   & v & v & o \\\hline
Experiment tracking                  & v & v & v & v & v & v & v & v \\\hline
Pipeline                             & v & v &   & v &   &   &   & $\triangle$ \\\hline
Built-in pipeline component          & v &   &   &   &   &   &   & $\triangle$ \\\hline
Model management                     &   &   &   &   & v &   &   & o \\\hline
Model serving                        &   & v &   &   & v &   & v & $\triangle$ \\\hline
End-to-end platform                  &   & v &   &   &   &   &   & $\triangle$ \\\hline

\end{tabular}
\label{table:mlp_comparison}
\end{center}
\end{table*}

\begin{table*}[htbp]
\vspace*{0.5cm}
\caption{Notations and symbols used in Table 1}
\begin{center}
\begin{tabular}{|p{3cm}|l|}
\hline
\textbf{Notation}&\textbf{Description} \\\hline
TFX & Abbreviation for TensorFlow Extended  \\\hline
KF & Abbreviation for Kubeflow  \\\hline
DT & Abbreviation for Determined  \\\hline
MF & Abbreviation for Metaflow  \\\hline
MLF & Abbreviation for MLflow  \\\hline
NNI & Abbreviation for Neural Network Intelligence  \\\hline
AML & Abbreviation for Angel-ML \\\hline
v & Existing feature \\\hline
o & In-progress feature \\\hline
$\triangle$ & Future work \\\hline
\end{tabular}
\label{table:notation}
\end{center}
\end{table*}

\subsection{Comparing Kubernetes and YARN}
Submarine supports both Kubernetes and Apache Hadoop YARN as our container orchestration frameworks. Kubernetes is the most popular container orchestrator in the world, and thus almost every open source machine learning platform supports it, including TFX, Google Kubeflow, Determined-AI Determined, and so on. On the other hand, there have been very few, if any, other open source machine learning platforms that support YARN. The reasons that Submarine needs to support YARN are explained as follows.

\subsubsection{Avoid data movement}
YARN is the de-facto resource scheduler for big data analytic workloads in large enterprises, such as LinkedIn, Microsoft \cite{GPUClusters} and Yahoo \cite{LargeScale}. With the support of YARN, deep learning workloads can be directly conducted on Hadoop clusters where most of their data is stored, and thus avoid unnecessary data movement between Hadoop clusters and disaggregated deep learning clusters.

\subsubsection{Support Hadoop ecosystem}
YARN is a main component of Apache Hadoop. With the support of YARN, Submarine easily integrates with projects in the Hadoop ecosystem. For instance, Submarine integrates Azkaban \cite{Azkaban} which allows data scientists to submit a set of workflow tasks with Spark for data preprocessing and TensorFlow for distributed deep learning directly to Azkaban from Zeppelin notebooks \cite{Zeppelin}.

\subsubsection{Support fine-grained GPU scheduling}
Most clusters are heterogeneous in the real world. To improve resource utilization of heterogeneous clusters, YARN supports different compute resources such as memory, CPU, GPU, and FPGA. GPU scheduling is especially important for efficiency of distributed deep learning workloads.

\hspace{5pt} Typically, distributed deep learning workloads require gang scheduling reducing the flexibility of scheduling, so GPU scheduling is a key factor in efficiency of distributed deep learning workloads. Due to the effect of locality on GPU utilization described in \cite{GPUClusters}, a locality-aware GPU scheduler can improve GPU utilization significantly via reducing resource fragmentation and synchronization overheads. YARN provides GPU topology scheduling \cite{YARN-8851}, a locality-aware GPU scheduling strategy. Nevertheless, as described in \cite{AlibabaTalk}, Kubernetes scheduler does not provide a native fine-grained GPU scheduler, and thus users need to develop customized solutions by themselves, such as gpushare-scheduler-extender in Alibaba Cloud \cite{Aliyun}. In addition to GPU, YARN also supports fine-grained CPU, memory, and FPGA scheduling.

\subsubsection{Focus on data-intensive applications}
The design goals of YARN and Kubernetes are quite different. YARN can schedule more than 1000 containers per second, but Kubernetes can only schedule about 100 containers per second due to latency. Kubernetes stores plenty of data in etcd \cite{etcd} which causes long latency, and thus the scheduling performance is limited. Unlike Kubernetes, YARN only persists application-level metadata, and thus the latency is short. As the result, YARN is more suitable for data-intensive applications, whereas Kubernetes is optimized for long-running applications. In other words, if users transplant data-intensive applications from Apache YARN to Kubernetes, the efficiency will decrease significantly.  To conclude, Kubernetes lies at the heart of DevOps, and YARN plays a key role in data engineering.

\subsubsection{Support hierarchy queue}
YARN natively supports the hierarchical queue which is helpful for multi-tenant support and cluster utilization. On the other hand, Kubernetes needs additional third-party schedulers like YuniKorn \cite{yunikorn} to improve the scheduling performance and provide the hierarchical queue ability.

\subsection{Portability}
Some machine learning platforms strongly depend on the proprietary services provided by public cloud providers. For instance, Metaflow leverages AWS Batch Service to execute pipelines in a distributed manner. Hence, it is not suitable for companies with extensive data center footprint. On the contrary, Submarine can be deployed on on-premise clusters managed by Kubernetes and YARN. Submarine can be deployed in the public cloud, too, because all major cloud vendors support Kubernetes, including Amazon Elastic Kubernetes Service (EKS) in AWS, Google Kubernetes Engine (GKE) in GCP, and Azure Kubernetes Service (AKS) in Microsoft Azure. 

\subsection{Supporting multiple machine learning frameworks}
The tight coupling between the modeling layer and the infrastructure layer, caused by the machine learning platform designer’s desire to support a single machine learning framework, often forces the developers to resort to “glue code”, which is the auxiliary code to connect the machine learning code with the rest of the stack, inevitably limiting the flexibility and the extensibility of the system. The machine learning framework landscape is rapidly evolving. From a future-proof point of view, it is best to make the framework interchangeable. To avoid glue code, Submarine provides a common interface that supports popular machine learning frameworks, including TensorFlow, PyTorch, and MXNet. Data scientists are able to try and identify the most appropriate framework without significantly refactor the code. 

\subsection{High-level training SDK and Predefined Template Service}
Machine learning techniques have been widely used in various fields such as finance, medicine, and education. However, most experts in these fields are not familiar with machine learning techniques, and thus they spend too much time on learning machine learning framework API rather than on solving domain-specific problems.

\hspace{5pt} Submarine creates high-level Python SDK and predefined template to address the challenge. Submarine Python SDK provides many easy-to-use and domain-specific machine learning APIs. With the help of Submarine Python SDK, users can implement models in a few lines of Python code and focus on domain-specific problems. Moreover, Submarine Predefined Template Service enables users to run experiments without writing any code.

\subsection{End-to-end platform}
We define an end-to-end platform as follows. First, the platforms need to support each stage in the model lifecycle, including data preparation, prototyping, model training and model serving. Second, the platforms must support containerization and be able to be deployed on resource orchestrator clusters. In other words, users can manage their models in customizable containers which can be deployed, scaled, and managed by resource orchestrators. An end-to-end platform allows data scientists to create end-to-end machine learning workflows without switching toolsets frequently and manage the whole machine learning lifecycle in a consistent user interface. Third, end-to-end platforms must be capable to manage the whole model lifecycle in a consistent user interface.

\subsection{Learning curve}
Existing machine learning platforms requires deep expertise to use and operate. In MLflow, data scientists need to write a Kubernetes Job Spec to run an MLflow project on Kubernetes. To solve this problem, Submarine enables data scientists to run distributed training jobs on Kubernetes or YARN easily via a user-friendly workbench UI. In Determined, data scientists need to rewrite their model codes to integrate with its API. On the other hand, data scientists can execute their code on Submarine without modification. 

\section{Use cases \label{section:use_cases}}
Submarine has been widely used in many technology companies, including Ke.com and LinkedIn. We summarized the main points of these use cases as follows.

\subsection{Ke.com}
Ke.com, also known as Beike in Chinese, is the leading integrated online and offline platform for housing transactions and services in China. Data scientists develop models for speech recognition to improve customer service quality. They run machine learning workloads on a 30+ node Submarine cluster, in which each node has 2 GPUs. The performances of these speech recognition workloads running on two nodes can achieve 1.8 times faster than running on a single node.

\subsection{LinkedIn}
The mission of LinkedIn is to connect the world’s professionals to make them more productive and successful. LinkedIn is the world's largest professional network with 722+ million members in more than 200 countries and territories worldwide. It has a 50+ node Submarine cluster in which each node is equipped with 5 GPUs. Submarine is primarily applied to speed up the model training of the BERT-Large model \cite{devlin2018bert}. This model for recommendation system has 24 layers and 300+ million parameters to predict and understand user behaviors automatically. In addition, more than 3500 experiments run in the Submarine cluster per day.

\section{Future work \label{section:future_work}}
In the future, the Submarine community plans to focus on three features. Firstly, a feature store which enables data scientists to reuse features instead of rebuilding these features repeatedly. Secondly, a workflow pipeline management system which enables users to build and manage reusable machine learning workflow. Additionally, built-in pipeline components allow users to build their pipelines easily. Lastly, a containerized model serving service which enables users to serve machine learning models written by different machine learning frameworks.

\section{Conclusions \label{section:conclusion}}
The ubiquity of machine learning in production imposes plenty of challenges on both infrastructure administrators and users including expert data scientists and citizen data scientists. However, existing platforms only take expert data scientists into consideration, and thus fail to meet the needs of both infrastructure administrators and users. Unlike existing platforms, Submarine provides a unified solution for infrastructure administrators, expert data scientists, and citizen data scientists. In other words, Submarine is a flexible and portable platform for infrastructure administrators, a unified system to streamline the workflow from idea to serving models in production, and an easy-to-use platform for citizen data scientists to save their time on machine learning framework API and help them focus on domain-specific problems. In conclusion, the design principle of Submarine is best explained by Alan Kay’s quote “Make simple things simple, and complex things possible.”

\begin{acks}
The authors would like to thank the Apache Submarine community for the discussions around the design, and also acknowledge the contributions of the following people: https://github.com/apache/submarine/graphs/contributors. 

\hspace{5pt} Hung-Chang Hsiao was partially supported by the Intelligent Manufacturing Research Center (iMRC) from The Featured Areas Research Center Program within the framework of the Higher Education Sprout Project by the Ministry of Education (MOE) and Ministry of Science and Technology (MOST) under Grant MOST 108-2218-E-006-029 in Taiwan.
\end{acks}

\bibliographystyle{ACM-Reference-Format}
\bibliography{sample-base}


\begin{thebibliography}{45}


\ifx \showCODEN    \undefined \def \showCODEN     #1{\unskip}     \fi
\ifx \showDOI      \undefined \def \showDOI       #1{#1}\fi
\ifx \showISBNx    \undefined \def \showISBNx     #1{\unskip}     \fi
\ifx \showISBNxiii \undefined \def \showISBNxiii  #1{\unskip}     \fi
\ifx \showISSN     \undefined \def \showISSN      #1{\unskip}     \fi
\ifx \showLCCN     \undefined \def \showLCCN      #1{\unskip}     \fi
\ifx \shownote     \undefined \def \shownote      #1{#1}          \fi
\ifx \showarticletitle \undefined \def \showarticletitle #1{#1}   \fi
\ifx \showURL      \undefined \def \showURL       {\relax}        \fi
\providecommand\bibfield[2]{#2}
\providecommand\bibinfo[2]{#2}
\providecommand\natexlab[1]{#1}
\providecommand\showeprint[2][]{arXiv:#2}

\bibitem[\protect\citeauthoryear{??}{Lar}{2015}]%
        {LargeScale}
 \bibinfo{year}{2015}\natexlab{}.
\newblock \bibinfo{title}{Large Scale Distributed Deep Learning on Hadoop
  Clusters}.
\newblock
\newblock
\urldef\tempurl%
\url{https://developer.yahoo.com/blogs/129872361846/}
\showURL{%
\tempurl}


\bibitem[\protect\citeauthoryear{??}{YAR}{2018}]%
        {YARN-8851}
 \bibinfo{year}{2018}\natexlab{}.
\newblock \bibinfo{title}{[YARN-8851] GPU hierarchy/topology scheduling support
  based on pluggable device framework}.
\newblock
\newblock
\urldef\tempurl%
\url{https://issues.apache.org/jira/browse/YARN-8821}
\showURL{%
\tempurl}


\bibitem[\protect\citeauthoryear{??}{yun}{2020}]%
        {yunikorn}
 \bibinfo{year}{2020}\natexlab{}.
\newblock \bibinfo{booktitle}{\emph{Apache YuniKorn (Incubating)}}.
\newblock
\urldef\tempurl%
\url{https://yunikorn.apache.org/}
\showURL{%
Retrieved November 21, 2020 from \tempurl}


\bibitem[\protect\citeauthoryear{??}{Azk}{2020}]%
        {Azkaban}
 \bibinfo{year}{2020}\natexlab{}.
\newblock \bibinfo{title}{Azkaban}.
\newblock
\newblock
\urldef\tempurl%
\url{https://github.com/azkaban/azkaban}
\showURL{%
\tempurl}


\bibitem[\protect\citeauthoryear{??}{Swa}{2020}]%
        {Swarm}
 \bibinfo{year}{2020}\natexlab{}.
\newblock \bibinfo{title}{Docker Swarm}.
\newblock
\newblock
\urldef\tempurl%
\url{https://docs.docker.com/engine/reference/commandline/swarm/}
\showURL{%
\tempurl}


\bibitem[\protect\citeauthoryear{??}{etc}{2020}]%
        {etcd}
 \bibinfo{year}{2020}\natexlab{}.
\newblock \bibinfo{title}{etcd}.
\newblock
\newblock
\urldef\tempurl%
\url{https://github.com/etcd-io/etcd}
\showURL{%
\tempurl}


\bibitem[\protect\citeauthoryear{??}{NNI}{2020}]%
        {NNI}
 \bibinfo{year}{2020}\natexlab{}.
\newblock \bibinfo{title}{Microsoft NNI}.
\newblock
\newblock
\urldef\tempurl%
\url{https://github.com/microsoft/nni}
\showURL{%
\tempurl}


\bibitem[\protect\citeauthoryear{Abadi, Barham, Chen, Chen, Davis, Dean, Devin,
  Ghemawat, Irving, Isard, Kudlur, Levenberg, Monga, Moore, Murray, Steiner,
  Tucker, Vasudevan, Warden, Wicke, Yu, and Zheng}{Abadi et~al\mbox{.}}{2016}]%
        {Tensorflow}
\bibfield{author}{\bibinfo{person}{M. Abadi}, \bibinfo{person}{P. Barham},
  \bibinfo{person}{J. Chen}, \bibinfo{person}{Z. Chen}, \bibinfo{person}{A.
  Davis}, \bibinfo{person}{J. Dean}, \bibinfo{person}{M. Devin},
  \bibinfo{person}{S. Ghemawat}, \bibinfo{person}{G. Irving},
  \bibinfo{person}{M. Isard}, \bibinfo{person}{M. Kudlur}, \bibinfo{person}{J.
  Levenberg}, \bibinfo{person}{R. Monga}, \bibinfo{person}{S. Moore},
  \bibinfo{person}{D.G. Murray}, \bibinfo{person}{B. Steiner},
  \bibinfo{person}{P. Tucker}, \bibinfo{person}{V. Vasudevan},
  \bibinfo{person}{P. Warden}, \bibinfo{person}{M. Wicke}, \bibinfo{person}{Y.
  Yu}, {and} \bibinfo{person}{X. Zheng}.} \bibinfo{year}{2016}\natexlab{}.
\newblock \showarticletitle{TensorFlow: A System for Large-Scale Machine
  Learning}. In \bibinfo{booktitle}{\emph{Proceedings of the 12th USENIX
  Conference on Operating Systems Design and Implementation}} (Savannah, GA,
  USA) \emph{(\bibinfo{series}{OSDI'16})}. \bibinfo{publisher}{USENIX
  Association}, \bibinfo{address}{USA}, \bibinfo{pages}{265–283}.
\newblock
\showISBNx{9781931971331}


\bibitem[\protect\citeauthoryear{Airflow}{Airflow}{2020}]%
        {Airflow}
Airflow \bibinfo{year}{2020}\natexlab{}.
\newblock \bibinfo{title}{Apache Airflow}.
\newblock
\newblock
\urldef\tempurl%
\url{https://airflow.apache.org/}
\showURL{%
\tempurl}


\bibitem[\protect\citeauthoryear{Aliyun}{Aliyun}{2020}]%
        {Aliyun}
Aliyun \bibinfo{year}{2020}\natexlab{}.
\newblock \bibinfo{title}{AliyunContainerService gpushare-scheduler-extender}.
\newblock
\newblock
\urldef\tempurl%
\url{https://github.com/AliyunContainerService/gpushare-scheduler-extender}
\showURL{%
\tempurl}


\bibitem[\protect\citeauthoryear{Angelml}{Angelml}{2020}]%
        {Angelml}
Angelml \bibinfo{year}{2020}\natexlab{}.
\newblock \bibinfo{title}{Tencent Angel-ML}.
\newblock
\newblock
\urldef\tempurl%
\url{https://angelml.ai/}
\showURL{%
\tempurl}


\bibitem[\protect\citeauthoryear{AzureML}{AzureML}{2020}]%
        {AzureML}
AzureML \bibinfo{year}{2020}\natexlab{}.
\newblock \bibinfo{title}{Azure: Build, train, and deploy models from the cloud
  to the edge}.
\newblock
\newblock
\urldef\tempurl%
\url{https://azure.microsoft.com/en-us/services/machine-learning/}
\showURL{%
\tempurl}


\bibitem[\protect\citeauthoryear{Baer and Ngahane}{Baer and Ngahane}{2019}]%
        {SpotifyMachineLearningInfrastructure}
\bibfield{author}{\bibinfo{person}{J. Baer} {and} \bibinfo{person}{S.
  Ngahane}.} \bibinfo{year}{2019}\natexlab{}.
\newblock \bibinfo{booktitle}{\emph{The Winding Road to Better Machine Learning
  Infrastructure Through Tensorflow Extended and Kubeflow}}.
\newblock
\urldef\tempurl%
\url{https://engineering.atspotify.com/2019/12/13/the-winding-road-to-better-machine-learning-infrastructure-through-tensorflow-extended-and-kubeflow/}
\showURL{%
Retrieved November 21, 2020 from \tempurl}


\bibitem[\protect\citeauthoryear{Beam}{Beam}{2020}]%
        {Beam}
Beam \bibinfo{year}{2020}\natexlab{}.
\newblock \bibinfo{title}{Apache Beam}.
\newblock
\newblock
\urldef\tempurl%
\url{https://beam.apache.org/}
\showURL{%
\tempurl}


\bibitem[\protect\citeauthoryear{Brumbaugh, Kale, Luque, Nooraei, Park,
  Puttaswamy, Schiller, Shapiro, Shi, Siegel, Simha, Bhushan, Sbrocca, Yao,
  Yoon, Zanoyan, Zeng, Zhu, Cheong, Du, Feng, Handel, Hoh, Hone, and
  Hunter}{Brumbaugh et~al\mbox{.}}{2019}]%
        {Bighead}
\bibfield{author}{\bibinfo{person}{E. Brumbaugh}, \bibinfo{person}{A. Kale},
  \bibinfo{person}{A. Luque}, \bibinfo{person}{B. Nooraei}, \bibinfo{person}{J.
  Park}, \bibinfo{person}{K. Puttaswamy}, \bibinfo{person}{K.H. Schiller},
  \bibinfo{person}{E. Shapiro}, \bibinfo{person}{C. Shi}, \bibinfo{person}{A.N.
  Siegel}, \bibinfo{person}{N. Simha}, \bibinfo{person}{M. Bhushan},
  \bibinfo{person}{M. Sbrocca}, \bibinfo{person}{S.-J. Yao},
  \bibinfo{person}{P. Yoon}, \bibinfo{person}{V. Zanoyan}, \bibinfo{person}{X.
  Zeng}, \bibinfo{person}{Q. Zhu}, \bibinfo{person}{A. Cheong},
  \bibinfo{person}{M.G.-Q. Du}, \bibinfo{person}{J. Feng}, \bibinfo{person}{N.
  Handel}, \bibinfo{person}{A.K. Hoh}, \bibinfo{person}{J. Hone}, {and}
  \bibinfo{person}{B. Hunter}.} \bibinfo{year}{2019}\natexlab{}.
\newblock \showarticletitle{Bighead: A Framework-Agnostic, End-to-End Machine
  Learning Platform}.
\newblock \bibinfo{journal}{\emph{2019 IEEE International Conference on Data
  Science and Advanced Analytics (DSAA)}} (\bibinfo{year}{2019}),
  \bibinfo{pages}{551--560}.
\newblock


\bibitem[\protect\citeauthoryear{Chen, Li, Li, Lin, Wang, Wang, Xiao, Xu,
  Zhang, and Zhang}{Chen et~al\mbox{.}}{2015}]%
        {MXNet}
\bibfield{author}{\bibinfo{person}{T. Chen}, \bibinfo{person}{M. Li},
  \bibinfo{person}{Y. Li}, \bibinfo{person}{M. Lin}, \bibinfo{person}{N. Wang},
  \bibinfo{person}{M. Wang}, \bibinfo{person}{T. Xiao}, \bibinfo{person}{B.
  Xu}, \bibinfo{person}{C. Zhang}, {and} \bibinfo{person}{Z. Zhang}.}
  \bibinfo{year}{2015}\natexlab{}.
\newblock \showarticletitle{MXNet: A Flexible and Efficient Machine Learning
  Library for Heterogeneous Distributed Systems.}
\newblock \bibinfo{journal}{\emph{CoRR}}  \bibinfo{volume}{abs/1512.01274}
  (\bibinfo{year}{2015}).
\newblock
\urldef\tempurl%
\url{http://dblp.uni-trier.de/db/journals/corr/corr1512.html#ChenLLLWWXXZZ15}
\showURL{%
\tempurl}


\bibitem[\protect\citeauthoryear{Cloudera}{Cloudera}{2020}]%
        {Cloudera}
Cloudera \bibinfo{year}{2020}\natexlab{}.
\newblock \bibinfo{title}{Cloudera data science workbench: Self-service data
  science for the enterprise}.
\newblock
\newblock
\urldef\tempurl%
\url{https://www.cloudera.com/products/data-science-and-engineering/data-science-workbench.html/}
\showURL{%
\tempurl}


\bibitem[\protect\citeauthoryear{{D. Aronchick and J. Lewi}}{{D. Aronchick and
  J. Lewi}}{2017}]%
        {Kubeflow}
\bibfield{author}{\bibinfo{person}{{D. Aronchick and J. Lewi}}.}
  \bibinfo{year}{2017}\natexlab{}.
\newblock \bibinfo{title}{Introducing kubeflow - a composable, portable,
  scalable ml stack built for kubernetes}.
\newblock
\newblock
\urldef\tempurl%
\url{https://kubernetes.io/blog/2017/12/introducing-kubeflow-composable/}
\showURL{%
\tempurl}


\bibitem[\protect\citeauthoryear{{D. Baylor, E. Breck, H.-t. Cheng, N. Fiedel,
  C. Yu Foo, Z. Haque, S. Haykal, M. Ispir, V. Jain, L. Koc, C. Yuen Koo, L.
  Lew, C. Mewald, A. Naresh Modi, N. Polyzotis, S. Ramesh, S. Roy, S. Euijong
  Whang, M. Wicke, J. Wilkiewicz, X. Zhang, and M. Zinkevich}}{{D. Baylor, E.
  Breck, H.-t. Cheng, N. Fiedel, C. Yu Foo, Z. Haque, S. Haykal, M. Ispir, V.
  Jain, L. Koc, C. Yuen Koo, L. Lew, C. Mewald, A. Naresh Modi, N. Polyzotis,
  S. Ramesh, S. Roy, S. Euijong Whang, M. Wicke, J. Wilkiewicz, X. Zhang, and
  M. Zinkevich}}{2017}]%
        {TFX}
\bibfield{author}{\bibinfo{person}{{D. Baylor, E. Breck, H.-t. Cheng, N.
  Fiedel, C. Yu Foo, Z. Haque, S. Haykal, M. Ispir, V. Jain, L. Koc, C. Yuen
  Koo, L. Lew, C. Mewald, A. Naresh Modi, N. Polyzotis, S. Ramesh, S. Roy, S.
  Euijong Whang, M. Wicke, J. Wilkiewicz, X. Zhang, and M. Zinkevich}}.}
  \bibinfo{year}{2017}\natexlab{}.
\newblock \bibinfo{title}{TFX: A TensorFlow-Based Production-Scale Machine
  Learning Platform}.
\newblock , \bibinfo{numpages}{1387--1395}~pages.
\newblock
\urldef\tempurl%
\url{http://www.oreilly.com/webops-perf/free/kubernetes.csp}
\showURL{%
\tempurl}


\bibitem[\protect\citeauthoryear{{D. K. Rensin, Kubernetes}}{{D. K. Rensin,
  Kubernetes}}{2015}]%
        {Kubernetes}
\bibfield{author}{\bibinfo{person}{{D. K. Rensin, Kubernetes}}.}
  \bibinfo{year}{2015}\natexlab{}.
\newblock \bibinfo{title}{Kubernetes - Scheduling the Future at Cloud Scale}.
\newblock
\newblock
\urldef\tempurl%
\url{http://www.oreilly.com/webops-perf/free/kubernetes.csp}
\showURL{%
\tempurl}


\bibitem[\protect\citeauthoryear{Determined}{Determined}{2020}]%
        {Determined}
Determined \bibinfo{year}{2020}\natexlab{}.
\newblock \bibinfo{title}{Determined-AI Determined}.
\newblock
\newblock
\urldef\tempurl%
\url{https://determined.ai/}
\showURL{%
\tempurl}


\bibitem[\protect\citeauthoryear{Devlin, Chang, Lee, and Toutanova}{Devlin
  et~al\mbox{.}}{2018}]%
        {devlin2018bert}
\bibfield{author}{\bibinfo{person}{J. Devlin}, \bibinfo{person}{M.-W. Chang},
  \bibinfo{person}{K. Lee}, {and} \bibinfo{person}{K. Toutanova}.}
  \bibinfo{year}{2018}\natexlab{}.
\newblock \showarticletitle{BERT: Pre-training of Deep Bidirectional
  Transformers for Language Understanding}.
\newblock \bibinfo{journal}{\emph{arXiv preprint arXiv:1810.04805}}
  (\bibinfo{year}{2018}).
\newblock


\bibitem[\protect\citeauthoryear{Dunn}{Dunn}{2016}]%
        {Fblearner}
\bibfield{author}{\bibinfo{person}{J. Dunn}.} \bibinfo{year}{2016}\natexlab{}.
\newblock \bibinfo{title}{Introducing fblearner flow: Facebook’s ai
  backbone}.
\newblock
\newblock
\urldef\tempurl%
\url{https://code.fb.com/core-data/introducing-fblearner-flow-facebook-s-aibackbone/}
\showURL{%
\tempurl}


\bibitem[\protect\citeauthoryear{Guo, Tang, Ye, Li, and He}{Guo
  et~al\mbox{.}}{2017}]%
        {DeepFM}
\bibfield{author}{\bibinfo{person}{H. Guo}, \bibinfo{person}{R. Tang},
  \bibinfo{person}{Y. Ye}, \bibinfo{person}{Z. Li}, {and} \bibinfo{person}{X.
  He}.} \bibinfo{year}{2017}\natexlab{}.
\newblock \showarticletitle{DeepFM: A Factorization-Machine Based Neural
  Network for CTR Prediction}. In \bibinfo{booktitle}{\emph{Proceedings of the
  26th International Joint Conference on Artificial Intelligence}} (Melbourne,
  Australia) \emph{(\bibinfo{series}{IJCAI'17})}. \bibinfo{publisher}{AAAI
  Press}, \bibinfo{pages}{1725–1731}.
\newblock
\showISBNx{9780999241103}


\bibitem[\protect\citeauthoryear{Hermann and Balso}{Hermann and Balso}{2017}]%
        {Michelangelo}
\bibfield{author}{\bibinfo{person}{J. Hermann} {and} \bibinfo{person}{M.D.
  Balso}.} \bibinfo{year}{2017}\natexlab{}.
\newblock \showarticletitle{Meet michelangelo: Uber’s machine learning
  platform}.
\newblock  (\bibinfo{year}{2017}).
\newblock
\urldef\tempurl%
\url{https://eng.uber.com/michelangelo}
\showURL{%
\tempurl}


\bibitem[\protect\citeauthoryear{Hsu, Hu, Hung, Suresh, and Zhang}{Hsu
  et~al\mbox{.}}{2019}]%
        {Tony}
\bibfield{author}{\bibinfo{person}{A Hsu}, \bibinfo{person}{K Hu},
  \bibinfo{person}{J Hung}, \bibinfo{person}{A Suresh}, {and}
  \bibinfo{person}{Z Zhang}.} \bibinfo{year}{2019}\natexlab{}.
\newblock \bibinfo{title}{TonY: An Orchestrator for Distributed Machine
  Learning Jobs}.
\newblock
\newblock
\showeprint[arxiv]{1904.01631}~[cs.DC]


\bibitem[\protect\citeauthoryear{Idoine}{Idoine}{2018}]%
        {CitizenDataScientist}
\bibfield{author}{\bibinfo{person}{C. Idoine}.}
  \bibinfo{year}{2018}\natexlab{}.
\newblock \bibinfo{booktitle}{\emph{Citizen Data Scientists and Why They
  Matter}}.
\newblock
\urldef\tempurl%
\url{https://blogs.gartner.com/carlie-idoine/2018/05/13/citizen-data-scientists-and-why-they-matter/}
\showURL{%
Retrieved November 21, 2020 from \tempurl}


\bibitem[\protect\citeauthoryear{Jeon, Venkataraman, Phanishayee, Qian, Xiao,
  and Yang}{Jeon et~al\mbox{.}}{2019}]%
        {GPUClusters}
\bibfield{author}{\bibinfo{person}{M. Jeon}, \bibinfo{person}{S. Venkataraman},
  \bibinfo{person}{A. Phanishayee}, \bibinfo{person}{U. Qian},
  \bibinfo{person}{W. Xiao}, {and} \bibinfo{person}{F. Yang}.}
  \bibinfo{year}{2019}\natexlab{}.
\newblock \showarticletitle{Analysis of Large-Scale Multi-Tenant GPU Clusters
  for DNN Training Workloads}. In \bibinfo{booktitle}{\emph{Proceedings of the
  2019 USENIX Conference on Usenix Annual Technical Conference}} (Renton, WA,
  USA) \emph{(\bibinfo{series}{USENIX ATC '19})}. \bibinfo{publisher}{USENIX
  Association}, \bibinfo{address}{USA}, \bibinfo{pages}{947–960}.
\newblock
\showISBNx{9781939133038}


\bibitem[\protect\citeauthoryear{Jupyter}{Jupyter}{2020}]%
        {Jupyter}
Jupyter \bibinfo{year}{2020}\natexlab{}.
\newblock \bibinfo{title}{Jupyter notebook}.
\newblock
\newblock
\urldef\tempurl%
\url{https://jupyter.org/}
\showURL{%
\tempurl}


\bibitem[\protect\citeauthoryear{{K. Zhang and Y. Che}}{{K. Zhang and Y.
  Che}}{2019}]%
        {AlibabaTalk}
\bibfield{author}{\bibinfo{person}{{K. Zhang and Y. Che}}.}
  \bibinfo{year}{2019}\natexlab{}.
\newblock \bibinfo{title}{Minimizing GPU Cost for Your Deep Learning on
  Kubernetes}.
\newblock
\newblock
\urldef\tempurl%
\url{https://events19.lfasiallc.com/events/kubecon-cloudnativecon-china-2019/schedule-english/}
\showURL{%
\tempurl}


\bibitem[\protect\citeauthoryear{Krizhevsky}{Krizhevsky}{2014}]%
        {krizhevsky2014weird}
\bibfield{author}{\bibinfo{person}{A. Krizhevsky}.}
  \bibinfo{year}{2014}\natexlab{}.
\newblock \bibinfo{title}{One weird trick for parallelizing convolutional
  neural networks}.
\newblock
\newblock
\showeprint[arxiv]{1404.5997}~[cs.NE]


\bibitem[\protect\citeauthoryear{Krizhevsky, Sutskever, and Hinton}{Krizhevsky
  et~al\mbox{.}}{2012}]%
        {CNN}
\bibfield{author}{\bibinfo{person}{A. Krizhevsky}, \bibinfo{person}{I.
  Sutskever}, {and} \bibinfo{person}{G.E. Hinton}.}
  \bibinfo{year}{2012}\natexlab{}.
\newblock \showarticletitle{ImageNet Classification with Deep Convolutional
  Neural Networks}.
\newblock In \bibinfo{booktitle}{\emph{Advances in Neural Information
  Processing Systems 25}}, \bibfield{editor}{\bibinfo{person}{F.~Pereira},
  \bibinfo{person}{C.~J.~C. Burges}, \bibinfo{person}{L.~Bottou}, {and}
  \bibinfo{person}{K.~Q. Weinberger}} (Eds.). \bibinfo{publisher}{Curran
  Associates, Inc.}, \bibinfo{pages}{1097--1105}.
\newblock
\urldef\tempurl%
\url{http://papers.nips.cc/paper/4824-imagenet-classification-with-deep-convolutional-neural-networks.pdf}
\showURL{%
\tempurl}


\bibitem[\protect\citeauthoryear{LeCun, Cortes, and Burges}{LeCun
  et~al\mbox{.}}{2010}]%
        {lecun2010mnist}
\bibfield{author}{\bibinfo{person}{Y. LeCun}, \bibinfo{person}{C. Cortes},
  {and} \bibinfo{person}{C. Burges}.} \bibinfo{year}{2010}\natexlab{}.
\newblock \showarticletitle{MNIST handwritten digit database}.
\newblock \bibinfo{journal}{\emph{ATT Labs [Online]. Available:
  http://yann.lecun.com/exdb/mnist}}  \bibinfo{volume}{2}
  (\bibinfo{year}{2010}).
\newblock


\bibitem[\protect\citeauthoryear{Metaflow}{Metaflow}{2020}]%
        {Metaflow}
Metaflow \bibinfo{year}{2020}\natexlab{}.
\newblock \bibinfo{title}{Netflix metaflow}.
\newblock
\newblock
\urldef\tempurl%
\url{https://github.com/Netflix/metaflow}
\showURL{%
\tempurl}


\bibitem[\protect\citeauthoryear{Moritz, Nishihara, Wang, Tumanov, Liaw, Liang,
  Paul, Jordan, and Stoica}{Moritz et~al\mbox{.}}{2017}]%
        {Ray}
\bibfield{author}{\bibinfo{person}{Philipp Moritz}, \bibinfo{person}{Robert
  Nishihara}, \bibinfo{person}{Stephanie Wang}, \bibinfo{person}{Alexey
  Tumanov}, \bibinfo{person}{Richard Liaw}, \bibinfo{person}{Eric Liang},
  \bibinfo{person}{William Paul}, \bibinfo{person}{Michael~I. Jordan}, {and}
  \bibinfo{person}{Ion Stoica}.} \bibinfo{year}{2017}\natexlab{}.
\newblock \showarticletitle{Ray: {A} Distributed Framework for Emerging {AI}
  Applications}.
\newblock \bibinfo{journal}{\emph{CoRR}}  \bibinfo{volume}{abs/1712.05889}
  (\bibinfo{year}{2017}).
\newblock
\showeprint[arxiv]{1712.05889}
\urldef\tempurl%
\url{http://arxiv.org/abs/1712.05889}
\showURL{%
\tempurl}


\bibitem[\protect\citeauthoryear{Ngahane and Goodsell}{Ngahane and
  Goodsell}{2018}]%
        {TwitterCortex}
\bibfield{author}{\bibinfo{person}{S. Ngahane} {and} \bibinfo{person}{D.
  Goodsell}.} \bibinfo{year}{2018}\natexlab{}.
\newblock \bibinfo{booktitle}{\emph{Productionizing ML with workflows at
  Twitter}}.
\newblock
\urldef\tempurl%
\url{https://blog.twitter.com/engineering/en_us/topics/insights/2018/ml-workflows.html}
\showURL{%
Retrieved November 21, 2020 from \tempurl}


\bibitem[\protect\citeauthoryear{Paszke, Gross, Massa, Lerer, Bradbury, Chanan,
  Killeen, Lin, Gimelshei, Antiga, Desmaison, Kopf, Yang, DeVito, Raison,
  Tejani, Chilamkurthy, Steiner, Fang, Bai, and Chintala}{Paszke
  et~al\mbox{.}}{2019}]%
        {PyTorch}
\bibfield{author}{\bibinfo{person}{A. Paszke}, \bibinfo{person}{S. Gross},
  \bibinfo{person}{F. Massa}, \bibinfo{person}{A. Lerer}, \bibinfo{person}{J.
  Bradbury}, \bibinfo{person}{G. Chanan}, \bibinfo{person}{T. Killeen},
  \bibinfo{person}{Z. Lin}, \bibinfo{person}{N. Gimelshei}, \bibinfo{person}{L.
  Antiga}, \bibinfo{person}{A. Desmaison}, \bibinfo{person}{A. Kopf},
  \bibinfo{person}{E. Yang}, \bibinfo{person}{Z. DeVito}, \bibinfo{person}{M.
  Raison}, \bibinfo{person}{A. Tejani}, \bibinfo{person}{S. Chilamkurthy},
  \bibinfo{person}{B. Steiner}, \bibinfo{person}{L. Fang}, \bibinfo{person}{J.
  Bai}, {and} \bibinfo{person}{S. Chintala}.} \bibinfo{year}{2019}\natexlab{}.
\newblock \showarticletitle{PyTorch: An Imperative Style, High-Performance Deep
  Learning Library}. In \bibinfo{booktitle}{\emph{Advances in Neural
  Information Processing Systems}},
  \bibfield{editor}{\bibinfo{person}{H.~Wallach},
  \bibinfo{person}{H.~Larochelle}, \bibinfo{person}{A.~Beygelzimer},
  \bibinfo{person}{F.~d\textquotesingle Alch\'{e}-Buc},
  \bibinfo{person}{E.~Fox}, {and} \bibinfo{person}{R.~Garnett}} (Eds.),
  Vol.~\bibinfo{volume}{32}. \bibinfo{publisher}{Curran Associates, Inc.},
  \bibinfo{pages}{8026--8037}.
\newblock
\urldef\tempurl%
\url{https://proceedings.neurips.cc/paper/2019/file/bdbca288fee7f92f2bfa9f7012727740-Paper.pdf}
\showURL{%
\tempurl}


\bibitem[\protect\citeauthoryear{Sagemaker}{Sagemaker}{2020}]%
        {Sagemaker}
Sagemaker \bibinfo{year}{2020}\natexlab{}.
\newblock \bibinfo{title}{Amazon sagemaker: Machine learning for every
  developer and data scientist}.
\newblock
\newblock
\urldef\tempurl%
\url{https://aws.amazon.com/sagemaker/}
\showURL{%
\tempurl}


\bibitem[\protect\citeauthoryear{Sculley, G.Holt, D.Golovin, E.vydov,
  T.Phillips, D.Ebner, V.Chaudhary, M.Young, Crespo, and Dennison}{Sculley
  et~al\mbox{.}}{2015}]%
        {sculley2015hidden}
\bibfield{author}{\bibinfo{person}{D. Sculley}, \bibinfo{person}{G.Holt},
  \bibinfo{person}{D.Golovin}, \bibinfo{person}{E.vydov},
  \bibinfo{person}{T.Phillips}, \bibinfo{person}{D.Ebner},
  \bibinfo{person}{V.Chaudhary}, \bibinfo{person}{M.Young},
  \bibinfo{person}{J.-F. Crespo}, {and} \bibinfo{person}{D. Dennison}.}
  \bibinfo{year}{2015}\natexlab{}.
\newblock \showarticletitle{Hidden technical debt in machine learning systems}.
  In \bibinfo{booktitle}{\emph{Advances in neural information processing
  systems}}. \bibinfo{pages}{2503--2511}.
\newblock


\bibitem[\protect\citeauthoryear{Smola and Narayanamurthy}{Smola and
  Narayanamurthy}{2010}]%
        {smola2010architecture}
\bibfield{author}{\bibinfo{person}{A. Smola} {and} \bibinfo{person}{S.
  Narayanamurthy}.} \bibinfo{year}{2010}\natexlab{}.
\newblock \showarticletitle{An architecture for parallel topic models}.
\newblock \bibinfo{journal}{\emph{Proceedings of the VLDB Endowment}}
  \bibinfo{volume}{3}, \bibinfo{number}{1-2} (\bibinfo{year}{2010}),
  \bibinfo{pages}{703--710}.
\newblock


\bibitem[\protect\citeauthoryear{Valohai}{Valohai}{2020}]%
        {Valohai}
Valohai \bibinfo{year}{2020}\natexlab{}.
\newblock \bibinfo{title}{Valohai MLOps platform}.
\newblock
\newblock
\urldef\tempurl%
\url{https://valohai.com/}
\showURL{%
\tempurl}


\bibitem[\protect\citeauthoryear{VertexAI}{VertexAI}{2021}]%
        {VertexAI}
VertexAI \bibinfo{year}{2021}\natexlab{}.
\newblock \bibinfo{title}{Vertex AI}.
\newblock
\newblock
\urldef\tempurl%
\url{https://cloud.google.com/vertex-ai}
\showURL{%
\tempurl}


\bibitem[\protect\citeauthoryear{YARN}{YARN}{2020}]%
        {YARN}
YARN \bibinfo{year}{2020}\natexlab{}.
\newblock \bibinfo{title}{Apache Hadoop YARN}.
\newblock
\newblock
\urldef\tempurl%
\url{https://hadoop.apache.org/docs/current/hadoop-yarn/hadoop-yarn-site/YARN.html}
\showURL{%
\tempurl}


\bibitem[\protect\citeauthoryear{Zahariai}{Zahariai}{2018}]%
        {Mlflow}
\bibfield{author}{\bibinfo{person}{M. Zahariai}.}
  \bibinfo{year}{2018}\natexlab{}.
\newblock \bibinfo{title}{Introducing mlflow: an open source machine learning
  platform}.
\newblock
\newblock
\urldef\tempurl%
\url{https://databricks.com/blog/2018/06/05/introducing-mlflow-an-open-source-machine-learning-platform.html}
\showURL{%
\tempurl}


\bibitem[\protect\citeauthoryear{Zeppelin}{Zeppelin}{2020}]%
        {Zeppelin}
Zeppelin \bibinfo{year}{2020}\natexlab{}.
\newblock \bibinfo{title}{Zeppelin}.
\newblock
\newblock
\urldef\tempurl%
\url{https://zeppelin.apache.org/}
\showURL{%
\tempurl}


\end{thebibliography}

\appendix

\end{document}